 \documentclass[final,5p,times,twocolumn]{elsarticle}

\usepackage{amssymb}

\usepackage{graphicx}
\usepackage{longtable}

\usepackage{graphicx,psfrag}
\usepackage{bm}
\usepackage{mathbbol,verbatim}
\usepackage{booktabs}
\usepackage{multirow}
\usepackage{slashed}
\usepackage{graphics}
\usepackage{color,ulem}
\usepackage{multirow}
\usepackage{amsmath}
\usepackage{tikz}
\usepackage{soul}
\usetikzlibrary{trees}
\usetikzlibrary{decorations.pathmorphing}
\usetikzlibrary{decorations.markings}
\usetikzlibrary{decorations, decorations.markings, decorations.pathmorphing, arrows, graphs, shapes.geometric, snakes}
\usetikzlibrary{arrows}

\def\beq{\begin{equation}}
\def\eeq{\end{equation}}
\newcommand{\beqa}{\begin{eqnarray}} 
\newcommand{\eeqa}{\end{eqnarray}}
\newcommand{\barr}{\begin{array}}
\newcommand{\earr}{\end{array}}

\def\gs{\mathrel{
   \rlap{\raise 0.511ex \hbox{$>$}}{\lower 0.511ex \hbox{$\sim$}}}}
\def\ls{\mathrel{
   \rlap{\raise 0.511ex \hbox{$<$}}{\lower 0.511ex \hbox{$\sim$}}}}
   \def\beq{\begin{equation}}
\def\eeq{\end{equation}}
\def\bea{\begin{equation}}
\def\eea{\end{equation}}

\def\ga{\gamma}

\usepackage{cancel}

\def\lapp{\mathrel{\rlap{\raise.5ex\hbox{$<$}}
                    {\lower.5ex\hbox{$\sim$}}}}
\def\gapp{\mathrel{\rlap{\raise.5ex\hbox{$>$}}
                    {\lower.5ex\hbox{$\sim$}}}}

\usepackage{graphicx,psfrag}
\usepackage{bm}
\usepackage{mathbbol,verbatim}
\usepackage{booktabs}
\usepackage{multirow}
\usepackage{slashed}
\usepackage{graphics}
\usepackage{color,ulem}
\usepackage{multirow}
\usepackage{multicol}
\usepackage{tikz}
\usetikzlibrary{trees}
\usetikzlibrary{decorations.pathmorphing}
\usetikzlibrary{decorations.markings}
\usetikzlibrary{decorations, decorations.markings, decorations.pathmorphing, arrows, graphs, shapes.geometric, snakes}
\usetikzlibrary{arrows}
\def\dis{\displaystyle}
   \def\beq{\begin{equation}}
\def\eeq{\end{equation}}
\def\bea{\begin{equation}}
\def\eea{\end{equation}}

\def\ga{\gamma}

\def\be{\begin{equation}}
\def\ee{\end{equation}}
\def\bea{\begin{eqnarray}}
\def\eea{\end{eqnarray}}
\usepackage{cancel}

\def\lapp{\mathrel{\rlap{\raise.5ex\hbox{$<$}}
                    {\lower.5ex\hbox{$\sim$}}}}
\def\gapp{\mathrel{\rlap{\raise.5ex\hbox{$>$}}
                    {\lower.5ex\hbox{$\sim$}}}}

\newcommand{\me}{\mathrm{e}}

\tikzstyle{block} = [rectangle, draw, fill=blue!20, 
    text width=5em, text centered, rounded corners, minimum height=4em]

\pdfoutput=1

\begin{document}

\begin{frontmatter}



\title{Universal Extra Dimension models with gravity mediated decays after LHC Run II data}


\author[a,b]{Kirtiman Ghosh}
\ead{kirti.gh@gmail.com}
\author[c]{Durmus Karabacak}
\ead{durmuskarabacak@gmail.com}
\author[d]{S. Nandi}
\ead{s.nandi@okstate.edu}

\address[a]{Institute of Physics, Sachivalaya Marg, Sainik School Post, Bhubaneswar 751005, India}
\address[b]{Homi Bhabha National Institute, Training School Complex, Anushakti Nagar, Mumbai 400085, India}
\address[c]{Mugla Sitki Kocman University, Faculty of Technology, Department of Energy Systems Engineering, Mugla 48000, Turkey}
\address[d]{Department of Physics and Oklahoma Center for High Energy Physics,
Oklahoma State University, Stillwater, OK 74078-3072, USA, and Department of Physics and Astronomy, Rice University, Houston, Texas 77005}

\begin{abstract}

In the 'fat-brane' realization of Universal Extra Dimension (UED) models, the gravity mediated decays of Kaluza-Klein (KK) excitations of the Standard Model (SM) particles offer interesting collider signals. Colored level-1 KK-particles (quarks $q^1$ and/or gluons $g^{1}$) are pair-produced at the colliders due to conserved KK-parity. These particles, then, cascade decay into lighter level-1 KK-particle in association with one or more SM particles until producing lightest KK particle (LKP). The {g}ravity mediation allow{s} LKP to  decay into photon or $Z$-boson plus gravity excitation, hence resulting in di-photon/$ZZ$/$Z\gamma$ plus missing transverse energy signatures at collider {experiments}. Alternatively, pair-produced level-1 KK quarks/gluons may directly decay into the corresponding SM  quark/gluon and a gravity excitation resulting in di-jet plus missing transverse energy signal. The ATLAS Collaboration has recently communicated {the results for} di-photon and multi-jet plus missing transverse energy {searches} with $36.1$  inverse-femtobarn of integrated luminosity at $13$ TeV center-of-mass energy. No significant excess of events above the SM expectation  {was} observed in both searches. We constrain the 'fat-brane' UED model parameters, namely the fundamental Planck mass $M_{D}$ and the size of small extra dimensions $R$, in the light of above-mentioned ATLAS searches.
\end{abstract}





\end{frontmatter}


The extra dimensional models offer another perspective on the shortcomings of the Standard Model (SM) and predict new signals at the
current and future collider experiments. In the case of ADD \cite{ADD}
model, for instance, the SM particles are localized on 3-brane (4-dimensional  manifold) and only gravity is allowed to propagate into '$N$' number of large extra dimensions. The four-dimensional Planck mass, is then diluted by the volume of the extra dimensional space $V_{N} \sim r^{N} $, where $N$ and $r$ are the number and size of large extra dimensions, resulting in higher dimensional Planck mass around a few TeV and hence offering a solution to naturalness/hierarchy problem. The same problem is also addressed by RS \cite{RS} model through introduction of warped metric. {On the other hand, there are a class of models, known as Universal Extra Dimension (UED) models, wherein some or all of the SM fields can access small (TeV$^{-1}$) extra dimension(s) \cite{antoniadis1,acd}.}  {Such scenarios} do not offer solutions {to the naturalness/hierarchy problem} as elegant as ADD or RS  {does however, could lead to a new mechanism of supersymmetry breaking \cite{antoniadis1}, relax the upper limit of the lightest supersymmetric neutral Higgs mass\cite{relax}, interpret the Higgs as a quark composite leading to a electroweak symmetry breaking (EWSB) without a fundamental scalar or Yukawa interactions \cite{Arkani-Hamed:2000hv}, lower the unification scale down to a few TeVs \cite{dienes}, give a different perspective to the issue of fermion mass hierarchy \cite{Arkani-Hamed:1999dc}, provide a cosmologically viable candidate for dark matter \cite{darkued1,darkued2}, predict the number of fermion generations to be an integral multiple of three \cite{2UED_fg}, explain the long life time of proton \cite{2UED_proton} and give rise to interesting signatures at collider experiments \cite{collider_ED,ED_LHC,coll2ued}.} {Our concern here is a specific and particularly interesting version of UED scenario, known as 'fat-brane' realization of UED, where in addition to TeV$^{-1}$ size extra dimension(s) (accessible to all SM fields and the gravity), large ($\sim$ eV$^{-1}$ to keV$^{-1}$ size) extra dimension(s) (accessible only to the gravity) are introduced \cite{NPB550,PLB482,Macesanu16}.}

 In UED, the SM gauge {symmetry}  is preserved  {on a} $3+1+m$ dimension{al} {space-time manifold} with {$m$} small {(}$\sim \rm{TeV}^{-1}${)} extra dimensions {being} compactified on different geometries. {A}ll  the SM {fields} are allowed to propagate into compact extra dimensions {resulting into} tower{s} of extra particles, called the Kaluza-Klein (KK) particles{.}  {Each particle in a KK-tower is} identified by {an} integer $n${, known as} the  {KK-}number.  {Translational symmetry along the extra dimension(s) ensures the} conservation of KK-number{s}.  {However,} in order to obtain the  chiral structure of {the} SM,  {one needs to introduce a} $Z_2$ symmetry. {For example, in the minimal version of UED (mUED) there is only a single flat
extra dimension ($y$), compactified on an $S^1 /Z_2$ orbifold with
radius $R$ \cite{acd}.} The $Z_2$ symmetry breaks the translation invariance along the extra dimension. {As a result, KK-number conservation breaks down at loop-level, leaving behind only a conserved KK-parity, defined as $(-1)^n$.}  {This discrete symmetry has several interesting consequences. KK-parity allows only pair production of level-1 KK-particles at the colliders, prohibits KK-modes from affecting tree-level EW precision observables, allows a level-1 KK-particles to decay into a lighter level-1 KK-particles and hence, ensures the stability of the lightest KK-particle (LKP). Being strongly interacting, level-1 KK quarks and gluons are copiously pair-produced at hadron colliders giving rise to multiple jets, leptons in association with missing transverse energy\footnote{$E_T\!\!\!\!\!/~$ results from the stable weakly interacting lightest level-1 KK-particle which remains invisible in the detector.} ($E_T\!\!\!\!\!/~$) signatures \cite{ED_LHC}. }

{In this work, we are interested in a particular variant of UED model where mUED is embedded in a $(4 + N)$ dimensional bulk  \cite{NPB550,PLB482} with $N$ large ($\sim$ eV$^{-1}$ to keV$^{-1}$ size) extra dimensions being accessed only by the gravity. The name 'fat brane' realization of UED came from the fact that the single small extra dimension of mUED (accessible to both matter and gravity) can be viewed as the thickness of the SM 3-brane in the $(4 +N)$-dimensional bulk.}
 In this scenario, both the SM particles and graviton would have KK excitations with different masses resulting from different compactifications. {The gravity induced interactions do not respect KK-number or KK-parity conservation among the KK-excitations of the SM particles. For example, the gravity induced interactions allow the level-1 KK-excitations of the SM fields to decay directly into corresponding SM particles by radiating a gravity excitation and thus, LKP is no more a stable particle. This makes the collider signatures of this model drastically different from the signatures of mUED. For example, gravity mediated decays of LKP give rise to photon(s) and/or $Z$-boson(s) in the final state. }  {On the other hand, if}  the Gravity Mediated Decays (GMD)  dominate  {over the KK-number conserving decays (KKCD)}, the pair-produced {strongly interacting} level-1 KK particles directly decay to their SM  {partners} in association with a gravity excitation resulting in di-jet plus large $\slashed E_T$\footnote{Here, $\slashed E_T$ results from  the gravity excitations escaping detection.}  {signature}. {In this work, we have studied the collider phenomenology of 'fat brane' realization of mUED in the context of recent ATLAS searches for di-photon/multi-jets plus $\slashed E_T$ signatures with $36.1$ inverse-femtobarn of integrated luminosity data collected at $13$ TeV center-of-mass energy of proton-proton collisions.}

\section{{The Model}}
\label{sec:2}
Minimal UED {is characterized} by one small extra dimension $y$ compactified on $S_1/Z_2${-orbifold} with $\mathcal{O}\sim ~\rm{TeV}^{-1}$ size radius $R$. All SM particles are assumed to propagate into  {$y$}. {The orbifolding is crucial in generating chiral zero modes for fermions. Each component of a 5-dimensional field is either even or odd under the orbifold projection. After compactification, the effective 4-dimensional Lagrangian can be written in terms of the respective zero modes (only for fields which are even under orbifold projection) and the KK excitations. The zero mode fields are identified with the SM particles. For the details of KK-decomposition of the SM fields in 5-dimension on $S_1/Z_2$-orbifold and resulting effective 4-dimensional Lagrangian, we refer the interested reader to Ref.~\cite{acd}.} 

The {tree level} mass of any level-n {KK-}particle  is  given by $m_n^2 = m_0^2+(nR^{-1})^2$, where $m_0$ is the corresponding SM particle mass. {F}or a moderate size of $R^{-1}> 500$ GeV, the mUED mass spectra is quite degenerate. The degeneracy can be partially lifted if radiative corrections are taken into account. There are two types of corrections:  Bulk corrections {arise from the winding of the internal loop around the compactified direction \cite{radi_matchev}, and}  are finite {and  nonzero only for the gauge boson KK-excitations.}  {On the other hand,} boundary{/orbifold} corrections  {are} logarithmically diverg{ent}. {The process of orbifolding introduces a set of fixed points in the fifth direction (two in the case of $S^1/Z_2$ compactification). Boundary corrections are the counterterms of the total orbifold correction, with the finite parts being completely unknown, and depend on the details of the ultraviolet completion.}  {Minimal} UED  assume{s} that all boundary terms  vanish at  {cutoff} scale $\Lambda > R^{-1}$ and  {hence, the corrections from the boundary terms, at a renormalization scale $\mu$ are proportional to ${\rm ln}(\Lambda^2/\mu^2)$}. 

{The mixing between the KK-excitations of the neutral electroweak gauge bosons is analogous to their SM counterparts and the mass eigenstates
and eigenvalues of the KK `photons' and `$Z$' bosons are obtained by
diagonalizing the following mass squared matrix.  
\[
\left( \barr{cc} 
\dis
\frac{n^2}{R^2}+ \hat{\delta} m_{B^n}^2 + \frac{1}{4}g^2 v^2 
& \dis
\frac{1}{4}g g^\prime v^2 
\\[2ex] 
\dis 
\frac{1}{4}g g^{\prime} v^2 & \dis
\frac{n^2}{R^2}+ \hat{\delta} m_{W^n}^2 +\frac{1}{4}{g^{\prime}}^2 v^2 \earr
\right), 
\]
where, $\hat{\delta} m_{B^n}^2$ and $\hat{\delta} m_{W^n}^2$ are the total
one-loop correction (including both bulk and boundary
contributions) for $B^{(n)}_\mu$ and $W_\mu^{3(n)}$, respectively and $g$ and $g^{\prime}$ are the SM gauge coupling corresponding to $SU(2)_L$ and $U(1)_Y$, respectively. It is important to note that, the
extent of mixing for non-zero KK-modes is miniscule and is progressively smaller for the higher KK-modes. As a consequence, the  $Z^{1}$ and $\gamma^1$ are,  for all practical purposes, essentially  $W^1_{3\mu}$ and $B^1_{\mu}$. This has profound consequences in the gravity mediated decays of LKP which will be discussed in the following.}

\subsection{Fat-brane mUED scenario \& gravity matter interactions}\label{sec:3}
In the fat-brane scenario, the gravity is allowed to propagate into $N$ large extra dimensions which are then compactified on a $N$-dimensional torus $T^N$ with volume $V_N \sim r^N$ where $r$ is  {the} size {of the $N$ large extra dimensions}. The $4$D Planck mass $M_{Pl}$ can be derived from the fundamental $(4+N)$-dimensional Planck mass $M_{D}$  as:
\beq\label{eq:add}
M_{Pl}^{2} = M_{D}^{N+2}(r/2\pi)^N.
\eeq
 Assuming there are $N$ such large extra dimensions denoted by $x^5,....,x^{4+N}$ with a common size of $r\sim$eV$^{-1}$  and one small extra dimension denoted by $y=x^4$ with TeV$^{-1}$ size one can write down the interaction of SM fields and the graviton in the higher dimension as:
\beq
\label{leq:lag}
\mathcal{S}_{int} = \int dx^{4+N} \delta(x^5)~ ...~ \delta(x^{4+N}) \sqrt{-\hat{g}}~\mathcal{L}_{m},
\eeq
where, $\mathcal{L}_m$ is the Lagrangian density for SM fermions and gauge bosons and the Higgs. $\hat{g}$ is higher dimensional flat metric defined as $\hat{g}_{\hat{\mu}\hat{\nu}}=\hat{\eta}_{\hat{\mu}\hat{\nu}}+\hat{\kappa}\hat{h}_{\hat{\mu}\hat{\nu}}$ where $\hat{\kappa}^2 = 16\pi G^{(4+N)}$
and $G^{(4+N)}$ is the Newton's constant in $(4+N)$ dimension. $\hat{h}_{\hat{\mu}\hat{\nu}}$, being $(4+N)$ dimensional tensor, has three components: the graviton $h_{\mu\nu}$ ($4$ dimensional tensor), the gravi-photons $A_{\mu i}$ ($N$ vectors) and $N^2$ the gravi-scalars $\phi_{ij}$, and defined as:
\begin{equation}
\hat{h}_{\hat{\mu}\hat{\nu}} = \frac{1}{\sqrt{V_N}}\begin{pmatrix}
h_{\mu\nu}+n_{\mu\nu}\phi & A_{\mu i} \\
A_{\nu j} & 2\phi_{ij}\\
\end{pmatrix},
\end{equation}
where $\phi=\phi_{ii}$,  $\mu,\nu = 0, 1, 2, 3$, and $i,j=4, 5, 6, ... , 3+N$. Since the gravity propagates into finite large extra dimensions it has the following KK decompositions:
\begin{eqnarray}
h_{\mu\nu}(x,y) &=& \sum_{\vec{n}} h_{\mu\nu}^{\vec{n}}(x)~ e^{i\frac{2\pi \vec{n}.\vec{y}}{r}}, \nonumber\\
A_{\mu i} (x,y)&=& \sum_{\vec{n}} A_{\mu i}^{\vec{n}}(x)~ e^{i\frac{2\pi \vec{n}.\vec{y}}{r}},\nonumber\\
\phi_{ij}(x,y) &=& \sum_{\vec{n}} \phi_{ij}^{\vec{n}}(x)~ e^{i\frac{2\pi \vec{n}.\vec{y}}{r}}, 
\end{eqnarray}
where, $\vec{n} = \{n_1,n_2,...,n_N\}$.  $\vec{n}= 0$ and $\vec{n}\neq 0$ respectively correspond to massless graviton ($h_{\mu\nu}$), gravi-photons ($A_{\mu i}$), gravi-scalars ($\phi_{ij}$) and their higher level KK-states. The mass of n-level excited graviton, gravi-photon and gravi-scalars are characterized by the size of large extra dimension `$r$' and KK-number vector $\vec{n}$ and reads $m_n = 2\pi |\vec{n}|/r$.
At the leading order of $\hat{\kappa}$ Eq. \ref{leq:lag} reads, 
\begin{equation}
\mathcal{S}_{int} \supset -\hat{\kappa}/2 \int d^{4+N} \!\!x ~\delta
(x^5) ... \delta(x^{4+N}) \hat{h}^{\hat{\mu}\hat{\nu}}T_{\hat{\mu}\hat{\nu}},
\end{equation}
where, $T_{\mu\nu}$, being the energy-momentum tensor in ($4+N$)D, is defined as
\begin{equation}
T_{\hat{\mu}\hat{\nu}} = \bigg(-\hat{\eta}_{\hat{\mu}\hat{\nu}} + 2 \frac{\partial\mathcal{L}_m}{\partial \hat{g}^{\hat{\mu}\hat{\nu}}}\bigg)_{\hat{g}=\hat{\eta}}.
\end{equation}
Expanding the interaction action in its ($\mu\nu$), ($\mu 4$) and ($44$) components of matter tensor one obtains the following expression:
\begin{eqnarray}
\mathcal{S}_{int} = -\kappa /2 \int &d^4x& \int_{0}^{\pi R} dy \sum \limits_{\vec{n}} \bigg [ \bigg( h_{\mu\nu}^{\vec{n}} + \eta_{\mu\nu}\phi^{\vec{n}} \bigg) T^{\mu\nu}\nonumber\\
 &-& 2A_{\mu 4}^{\vec{n}} T_4^\mu +2 \phi_{44}^{\vec{n}} T_{44}       \bigg] \me^{\frac{i 2\pi n_4 y }{r}},
\end{eqnarray}
where $\kappa$ is the Newton's constant in $4D$, defined as $\kappa \equiv \sqrt{16 \pi G^{(4)}}=V_{N}^{-1/2}\hat{\kappa}$. With the expressions defined above one can derive the Feynman rules corresponding the Gravity-matter interactions. These rules can be found in Ref. \cite{Macesanu16}.
\subsection{Gravity Mediated Decays (GMD) of Level-1 KK particles}
\label{sec:4}
In this section, we would like to present relevant expressions used for calculating GMD widths of level-1 KK particles. In the framework of `fat brane' scenarios, the SM particles are only allowed to propagate into a small but universal extra dimension along the large extra dimension(s) to which only gravity can propagate.  {This} configuration of the brane in the bulk {violates translation invariance along the small extra-dimension and hence,} does respect neither KK-number {conservation} nor KK-parity. This  enables  KK particles {to decay directly into the corresponding SM particles in association with a gravity excitation}, namely, gravitons, gravi-vectors and gravi-scalars. {The total GMD width is given by,}  
\begin{equation}
\Gamma = \sum \limits_{\vec{n}} \Gamma_{\vec{n}} = \bigg [ \sum \limits  \Gamma_{h^{\vec{n}}} + \Gamma_{A^{\vec{n}}} + \Gamma_{\phi^{\vec{n}}} \bigg].
\end{equation}
{The gravity propagates in large extra dimensions and hence,} the mass splitting between KK-gravity  {excitations} are small, roughly $\Delta m = 2\pi/r \sim \rm{eV~ to~keV}$.  {The sum } {in the} above {equation}  {could} be replaced by integral {as follows}:
\begin{equation}
\sum\limits_{\vec{n}} \Gamma_{\vec{n}} \longrightarrow \int \Gamma_{\vec{n}}~ d^N\vec{n},
\end{equation}
where, $d^N\vec{n}$ represents the number of gravity excitation in a mass range ($m_{\vec{n}}, m_{\vec{n}}+ dm$). $\vec{n}^2$ is given by $m_{\vec{n}}^2/\Delta m^2$ since the level-$\vec{n}$ gravity excitation mass is $m_{\vec{n}}^2 = 4\pi^2 \vec{n}^2/r^2$. The number of gravity excitations in a mass range ($m_{\vec{n}}, m_{\vec{n}}+ dm$) is then given by the volume of annular space between two $N$-dimensional hypersphere with radii $m_{\vec{n}}/\Delta m$ and $(m_{\vec{n}}+dm)/\Delta m$:
\begin{equation}
d^{N}\vec{n} =(m_{\vec{n}}/\Delta m)^{N-1} \frac{d m}{\Delta m} d\Omega=\frac{1}{\Delta m^N}~ m_{\vec{n}}^{N-1} ~dm~d\Omega,
\end{equation}
where, $d\Omega$ is $N$-dimensional solid angle. Using Eq. \ref{eq:add} one can obtain $\Delta m^N= M_{D}^{N+2}/M_{Pl}^2$ and calculate the total GMD width by, 
\begin{equation}
\Gamma = \frac{M_{Pl}^2}{M_{D}^{N+2}} \int \Gamma_{\vec{n}} m_{\vec{n}}^{N-1}~dm~d\Omega.
\label{eq:sum}
\end{equation}
\begin{figure*}[t]
\centering
\includegraphics[scale=0.5, angle =0]{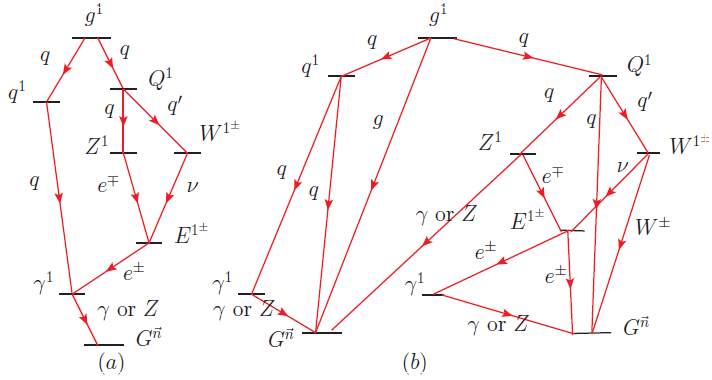}
\caption{ Decay cascade of level-1 gluon ($g^{1}$) (a) for $N=6$ and (b) for $N=2 \rm{~and~} 4$. $G^{\vec{n}} \ni h^{\vec{n}}, A^{\vec{n}},\phi^{\vec{n}}$. The level-1 neutrino, $\nu^{1}$, is omitted in sketch.}
\label{fig:decay_chain}
\end{figure*}

\section{Collider Phenomenology}
\label{sec:5}
In this section, we will discuss the phenomenology of level-1  {excitations of the SM fields} in the context of the LHC experiment. The particle spectrum of level-1 KK fields contains excited fermions ($SU(2)_L$-doublets: $Q^1$ and $L^1$; $SU(2)_L$-singlets $u^1$, $d^1$ and $e^1$), Higgses and gauge bosons (excited gluon: $g^1$, $W$-boson: $W^{1\pm}$ and $Z^1$ and photon: $\ga^1$). In the absence of electroweak symmetry breaking, the masses of all level-1 KK particles are given by $R^{-1}$. However, radiative corrections \cite{radi_matchev} remove  {this} degeneracy.  KK-fermions receive {positive} mass corrections from  {both} gauge interactions (with KK-gauge bosons) and Yukawa interactions.  The gauge fields receive mass corrections from the self-interactions and gauge interactions (with KK-fermions). Gauge interactions  give negative mass shift, while the self-interactions give positive mass shift. However, mass of the hypercharge gauge boson ($\ga^1$) receive only negative corrections from fermionic loops. Numerical computations show that the lightest KK-particle is the hypercharge gauge boson $\ga^1$ and the heaviest level-1 KK particle is the excited gluon ($g^1$). The radiative corrections are  {proportional to ${\rm ln}(\Lambda^2/\mu^2)$ where $\Lambda$ is the cutoff scale. The
perturbativity of the $U(1)_Y$ gauge coupling requires $\Lambda \leq 40R^{-1}$. However, much stronger bounds arise from the the running of the Higgs-boson self-coupling and the stability of the electroweak vacuum \cite{UED_VS2,Datta:2013xwa}.}  We choose  $\Lambda = 5 R^{-1}$ throughout this analysis. {The mass hierarchy between level-1 KK-particles after incorporating the radiative corrections is schematically shown in Fig.~\ref{fig:decay_chain}.}

Level-1 quarks and gluons{, being charged under $SU(3)_C$,} are abundantly {pair} produced at the LHC and their decay{s} give rise to interesting signatures. Before going into the details of {the signatures at the} LHC  {experiment}, it is important to discuss the decays of level-1 KK-particles. The decays of level-1 particles with emphasis on the gravity mediation was previously discussed in detail in Ref. \cite{Ghosh:2012zc}. For the sake of completeness of this article,  a brief discussion about the decays of the level-1 KK particles  {is presented} in the following:
\begin{figure*}[t]
\centering
\includegraphics[scale=.65, angle =-90]{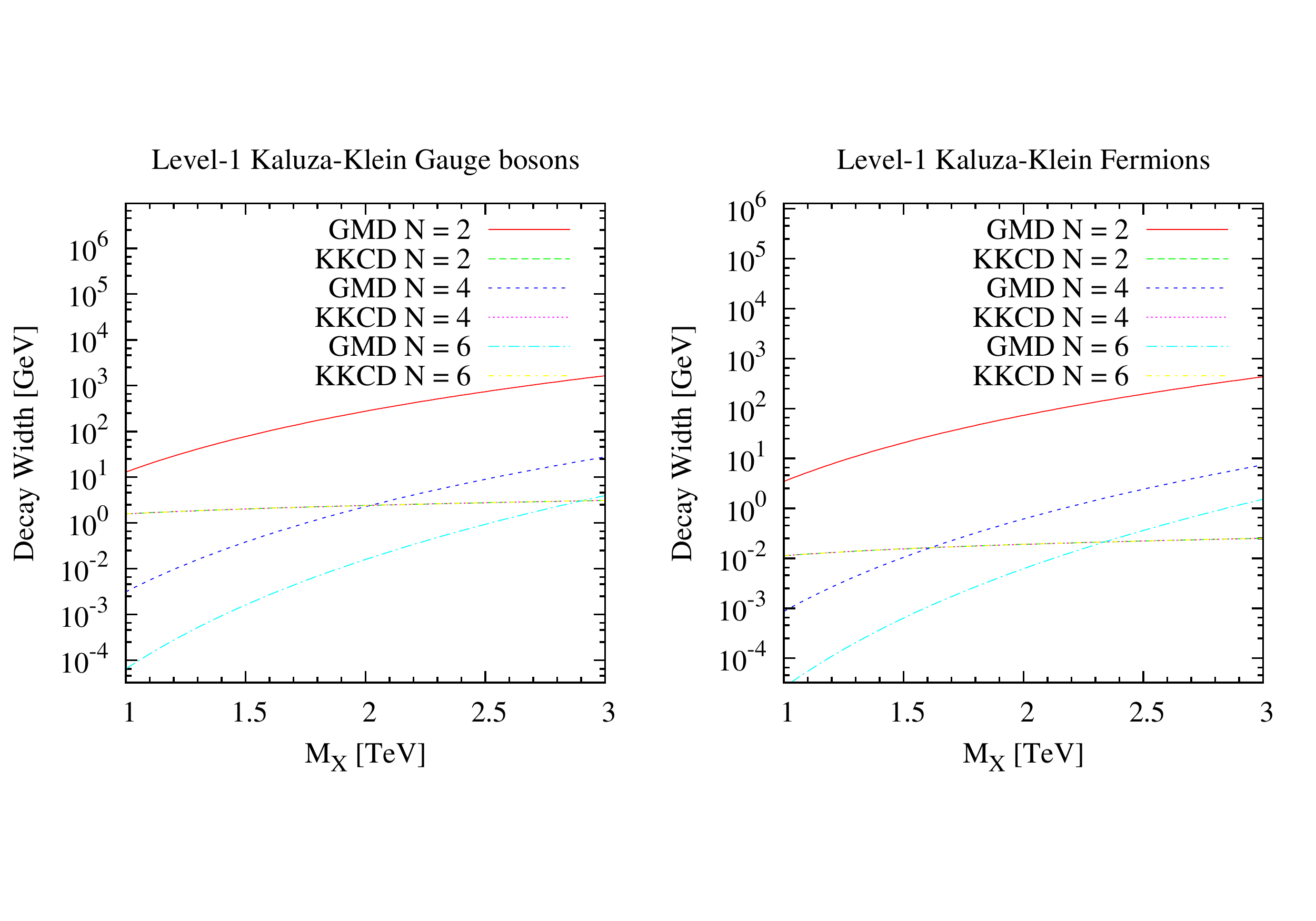}
\caption{ The KK-Number Conserving Decay (KKCD) and Gravity Mediated Decay (GMD) widths for level-1 KK gauge boson (left) and quarks (right) as a function of particle's mass $M_{X}$. $N$ is the number of large extra dimensions. $\Lambda R = 5$ and $M_D=5$ TeV are set in producing KKCD and GMD widths.}
\label{fig:widths}
\end{figure*}

\subsection{Decays}
In the framework of {'fat-brane' UED}, the decay mechanisms of KK particle can be categorized into: {\it KK-number Conserving Decay (KKCD)} and  {\it Gravity Mediated Decay (GMD)}. \\

\noindent{\it \uline{KK-number Conserving Decay (KKCD)}:} Conservation of KK-number (as well as KK-parity) ensures the decay of level-1 particles only into a lighter level-1 KK-particle in association with one or more SM particles. As a result, being the lightest level-1 KK particle, $\gamma^1$ does not have any KK-number conserving decay. For a fixed $R^{-1}$ and $\Lambda$, $g^1$, being the heaviest particle in the spectrum, can decay into doublet $Q^{1}$ and singlet ($u^1, d^1$) quarks with almost the same probability. The singlet quarks, in turn, can only decay into $\gamma^1$ and SM quark. On the other hand, the doublet quarks, can mostly decay into level-1 KK electroweak gauge bosons, namely $Z^1$ and $W^1$. The hadronic decays of $W^1$ and $Z^1$ are kinematically closed. Therefore, after decaying into level-1 KK leptons  and the corresponding SM lepton, they finally decay into SM leptons and $\gamma^1$. We also note that the masses and the KKCD widths of level-1 particles do not depend on the number of large extra dimensions, $N$ and  {are determined only by} the size of small extra dimension $R^{-1}$ and cutoff scale of the model $\Lambda$.   \\

\noindent{\it \uline{Gravity Mediated Decay (GMD)}:} As discussed in Section \ref{sec:3} and \ref{sec:4}, KK-number is not a conserved quantity for the gravity-matter interactions. Therefore, level-1 KK matter fields can decay into a level-$\vec{n}$ gravity excitation $G^{\vec{n}}$ ($G^{\vec{n}} \subset$ graviton, graviphoton, or graviscalar) and respective SM matter particles. The partial gravity mediated decay width of level-1 matter fields into a level-$\vec{n}$ gravity excitation $G^{\vec{n}}$ can be computed using the Feynman rules for the gravity-matter interactions and the total decay width is obtained by summing over all possible gravity excitations with mass smaller than the decaying particle as given in Eq.\ref{eq:sum} (for details see Ref. \cite{Ghosh:2012zc}).

In Fig.~\ref{fig:widths}, we present the partial KKCD and GMD widths of level-1 KK gauge boson ($g^1$) (left panel) and quarks ($u^1, d^1$)  (right panel) as a function of particle mass for $N = 2, 4,\rm{~and~} 6$. KKCD widths are essentially independent of the number of  large of extra dimensions $N$. On the other hand, the GMD widths are quite sensitive to the number of large extra dimensions and increases for decreasing value of $N$. This feature can be attributed to the fact that smaller $N$ (for example, $N=2$) corresponds to small mass splittings between KK-gravity excitations and hence, larger density of KK-gravity states and larger GMD widths. {Fig.~\ref{fig:widths} shows that}   KKCD and GMD widths are comparable for $N = 4$   {whereas,} GMD(KKCD) widths are  {larger} for $N = 2(6)$. {This has interesting consequences at the collider experiments which will be discussed in the following.}

\subsection{Collider Signatures}
{In Fig.~\ref{fig:decay_chain},} we  {schematically present} the decay cascade of level-1 KK gluon ($g^1$). In the left (right) panel of Fig~\ref{fig:decay_chain}, we show the dominant decay modes of $g^1$ for $N=6~ (2,4)$. As  {argued in the previous paragraph,}  for $N=6$,  KKCD  dominate{s}  {over GMD and hence}, $g^1$  {dominantly} decays to level-1 KK quarks (doublet ($Q^1$) or singlet ($q^1$) with almost equal probability) plus corresponding SM quark {followed by the KKCD of KK-quarks into lighter KK-particle in association with SM quarks}. The KK number conserving decay  {cascade} terminate{s} at the LKP ($\gamma^1$) since {the KKCD is forbidden for} the LKP. {However, in the frame of 'fat brane' UED, gravity-matter interactions allow LKP to}  decay into $\gamma \rm{~or~} Z$-boson plus a gravity excitation $G^{\vec{n}}$. Therefore, for $N=6$, {pair production followed by the subsequent cascade}  decay of level-1 quarks/gluons give {rise to} $\ga \ga,~\ga Z \rm{~or~} ZZ + X+\slashed{E_T}$ {final states at the hadron} collider  {experiments} where $X$ corresponds to the SM jets/leptons emitted in the KKCD cascade. $G^{\vec{n}}$ {remains invisible in the detector and hence, results into}  missing transverse energy signature. The picture  radically changes for $N=2\rm{~and~}4$ for which the dominant decay modes for level-1 KK-particles are shown in {the} right panel (b) of Fig.~\ref{fig:decay_chain}.  For $N=2$, the GMD width dominates over KKCD width for a particle mass $M_X\gtrsim 1$ TeV where $X=g^1, Q^1(q^1)$ (see Fig.~\ref{fig:widths}). Hence, $g^1(Q^1/q^1)$ dominantly decay into gluon(quark) plus a gravity excitation via gravity indused interactions. Therefore, for $N=2${, the pair production and subsequent}  decay of level-1 KK gluons/quarks  {give rise to} di-jet plus missing transverse energy  {signature}. The similar conclusion can also be drawn  {for} $N=4$ for $M_{g^1(q^1/Q^1)} \gtrsim 2 (1.7)$ TeV {where GMD dominates over the KKCD.}     

{After discussing the decays and hence, the signal topologies of level-1 KK particles in the framework of 'fat brane' UED, we are now equiped enough to discuss the impact of the LHC Run II data on the parameter space of the present model. In this work, we have studied dijet and di-photon + $\slashed E_T$ signatures in the context of recent LHC results which will be discussed briefly in the following.

\begin{table*}[t]
\begin{center}
\begin{tabular}{|c||c|c|c|c|c|c|c|}
\hline
\multirow{2}{*}{\textbf{Cuts}}&\multicolumn{7}{|c|}{\textbf{Signal Region}} \\
\cline{2-8}
 & \textbf{2j-1200} &\textbf{2j-1600} &\textbf{2j-2000} &\textbf{2j-2400}&\textbf{2j-2800} &\textbf{2j-3600} &\textbf{2j-2100} \\
\hline
$\slashed{E}_T$ [GeV]&\multicolumn{7}{c|}{250}\\
\hline
$p_T(j_1)$ [GeV] &250&300&\multicolumn{4}{|c|}{350}&600\\
\hline
$p_T(j_2)$ [GeV]&250&300&\multicolumn{4}{|c|}{350}&50\\
\hline
$|\eta(j_{12})|<$&0.8&\multicolumn{4}{|c|}{1.2}&\multicolumn{2}{|c|}{-}\\
\hline
$\Delta \phi$($jet_{1,2,(3)},\vec{\slashed E_T})_{min}>$ & \multicolumn{6}{|c|}{0.8}&0.4\\
\hline
$\Delta \phi (jet_{i>3},\vec{\slashed E_T})_{min}>$&\multicolumn{6}{|c|}{0.4}&0.2\\
\hline
$\slashed{E}_T/ \sqrt{H_T}>$[GeV$^{1/2}$]& 14&\multicolumn{5}{|c|}{18}&26\\
\hline
m$_{\rm{eff}}\rm{(incl.)}>$[TeV]&12&16&20&24&28&36&21\\
\hline
$\sigma_{BSM}$[fb]&3.6&1.00&0.42&0.30&0.32&0.20&2.0\\
\hline
\end{tabular}
\end{center}
\caption{Cuts and the signal regions used by the ATLAS Collaboration \cite{ATLAS_multijet13} in multi-jet search along with model independent observed 95\% C.L. upper limits on the BSM contributions ($\sigma_{BSM}$) for different SRs. $\Delta \phi(j,\vec{\slashed E_T})_{min}$ is defined as the minimum azimuthal separation between the jets and missing transverse momenta. $H_T$ is the sum of all jets $p_T$. $m_{eff}(\rm{incl.})$ is the sum of all jets with $p_T> 50$ GeV and $\slashed E_T$.}
\label{tb:multijet_cuts}
\end{table*}

\subsubsection{Dijet$+\slashed E_T$ search}

{Recently, the ATLAS collaboration \cite{ATLAS_multijet13} has performed a dedicated search for multijet($2-6$ jets)$+\slashed E_T$ signatures using $36.1$ fb$^{-1}$ integrated luminosity data of proton-proton collision at $\sqrt{s} = 13$ TeV. The search was designed to probe strongly interacting supersymmetric particles namely, squarks and gluinos. However,}  on the ground of consistency between experimental data and the SM prediction{,} model independent $95\%$ CL upper limits {are set} on the { visible cross-section $<\epsilon\sigma>^{95}_{obs}$ defined as the product of cross section, acceptance and efficiency ($\sigma \times A \times \epsilon$)}  for a{ny} new  {scenario} beyond the SM physics.  {In this work,  we now perform an analogous exercise for mUED with gravity mediated decays. As it has been already argued in the previous section that 'fat brane' UED dominantly gives rise to dijet$+\slashed E_T$ signature at the hadron colliders for $N=2~\rm{and}~4$, we restrict ourselves to the ATLAS results for dijet$+\slashed E_T$ searches only which will be discussed in the following.}

In  {ATLAS} analysis, jet candidates are reconstructed by anti-$k_T$ jet clustering algorithm \cite{antikt} with $0.4$ jet radius parameter $\Delta R$. {Only jets with $p_T > 20$ GeV and $|\eta| <$ 2.8 are considered for further analysis.} Electron (muon) candidates are required to have $p_T>7$ GeV and lie within $|\eta| < 2.47 (2.7)$ rapidity range. After jet and lepton identification, any jet candidate within a distance $\Delta R = \sqrt{(\Delta \eta)^2 + (\Delta \phi)^2} = 0.2$ of an electron is discarded. Moreover, if an electron (muon) and a jet are found  within $0.2 \le \Delta R < 0.4$($< \rm{min}(0.4,0.04+10 \rm{~GeV}/p_T^\mu)$), the object is interpreted as jet and the nearby electron (muon) candidate is removed. If a muon and jet are found within $\Delta R < 0.2$, then the object is interpreted as muon and the jet is discarded. Missing transverse energy calculation is based on all  {reconstructed} jets, leptons and all calorimeter clusters not associated to such objects. {Events with zero lepton and atleast one reconstructed jet with $p_T>50$ GeV are selected for further analysis.} The results of ATLAS multi-jet search is presented in  {different} inclusive Signal Regions (SRs) based on increasing number of jet {multiplicity} and tighter cut on $m_{eff}(incl.)$ which is defined as the scalar sum of all jet $p_T$'s with $p_T(jet)>50$ GeV and  $\slashed E_T$. {Here, we are only interested on ATLAS dijet searches. In Table}~\ref{tb:multijet_cuts}\footnote{The signal regions with higher jet multiplicities are omitted in the table since gravity mediated decays of KK-particles dominantly result into dijet signature and hence, for 'fat brane' UED, strongest exclusion limits come from the di-jet SRs.}, we present the cuts used by ATLAS {collaboration to define different di-jet SRs.} 

\subsubsection{Di-photon $+\slashed E_T$ search}
{In 'fat brane' UED scenario for $N=6$, pair productions of level-1 KK-quarks/gluons and their subsequent KK-number conserving cascade decay to $\gamma^1$ followed by the gravity mediated decay $\gamma^1 \to \gamma/Z+G^{\vec n}$ give rise to di-photon/$ZZ$/$\gamma Z$ $+\slashed E_T$ final states. These signatures are analogous to the signatures of gauge-mediated supersymmetry (GGM) breaking scenario where the decay of next-to-lightest supersymmetric particle (NLSP) to gravitino LSP in association with a photon gives rise to di-photon signature. With $36.1$ fb$^{-1}$ integrated luminosity data at $\sqrt s=13$ TeV, ATLAS collaboration \cite{ATLAS_di-photon13} have searched for di-photon$+\slashed E_T$ signature in the context of GGM model. We have used the model independent bounds on the visible di-photon$+\slashed E_T$ cross-section ($<\epsilon\sigma>^{95}_{obs}$) to constrain the parameter space of mUED with gravity mediated decays. The details of event selection for the ATLAS  di-photon$+\slashed E_T$ search can be found in Ref.~\cite{ATLAS_di-photon13} and also summarized in Table~\ref{tb:di-photon_cuts_new}. Reconstruction algorithms for jets, leptons\footnote{For  di-photon$+\slashed E_T$ search, jets with $p_T>30$ GeV and $|\eta|<2.8$ are considered. Whereas, Electron (muon) candidates are required to satisfy $p_T>25(25)$ GeV and $|\eta|<2.47(2.7)$ (excluding the transition region $1.37<|\eta| < 1.52$ between the barrel and endcap calorimeters).} and $\slashed E_T$ are analogous to the multijet analysis discussed in the previous section. The photon candidates are required to satisfy $p_T>25$ GeV and be in the range $|\eta|<2.37$ (excluding the transition region). Signal regions are classified into SR$_{S-L}^{\ga\ga}$ and SR$_{S-H}^{\gamma\gamma}$ to optimize the search for GGM scenarios with heavy and light gravitinos, respectively. The definition of SRs along with the ATLAS observed 95\% CL upper limits on BSM contribution to  di-photon$+\slashed E_T$ cross-sections are presented in Table~\ref{tb:di-photon_cuts_new}.}

\begin{table}
\begin{center}
\begin{tabular}{|c||c|c|}
\hline
\multirow{1}{*}{\textbf{Cuts}} & \textbf{$SR_{S-L}^{\gamma\gamma} $ } &\textbf{$SR_{S-H}^{\gamma\gamma} $ }\\
\hline
\hline
Number of photons & $\geq 2$& $\geq 2$ \\
\hline
$p_T(\ga_1)>$ [GeV]& 75& 75\\
\hline
$p_T(\ga_2)>$ [GeV]& 75& 75\\
\hline
$\slashed E_T>$ [GeV]& 150& 250\\
\hline
$H_T>$ [TeV] & 2.75 & 2.00\\
\hline
$\Delta \phi (\rm{jet},\slashed E_T)>$& 0.5 & 0.5\\
\hline
$\Delta \phi (\ga,\slashed E_T)>$&- & 0.5\\
\hline
\hline
$<\epsilon\sigma>_{\rm{obs}}^{95}$ [fb]& 0.083 & 0.083\\
\hline
\end{tabular}
\end{center}
\caption{Signal regions and cuts used by the ATLAS Collaboration \cite{ATLAS_di-photon13} in di-photon search along observed 95\% C.L. upper limit on model independent visible beyond the SM cross-section. $H_T$ is the scalar sum of the selected photons, any additional leptons and jets in the event. $\Delta \phi(\rm{jet},\slashed E_T)$ is the azimuthal separation between two leading jets with $p_T>75$ GeV and $\vec{\slashed E_T}$ vector. $\Delta \phi(\ga,\slashed E_T)$ is the azimuthal separation between selected photon and $\vec{\slashed E_T}$ vector. Visible transverse energy variable, $H_T$ is introduced as a sum of transverse energy of photons, any additional jets and leptons.}
\label{tb:di-photon_cuts_new}
\end{table}

\subsubsection{Event simulation \& object reconstruction}
{We used \textbf{PYTHIA} \cite{PYTHIA} with its mUED implementation \cite{ElKacimi:2009zj} to generate parton level events corresponding to pair productions of level-1 KK-quarks/gluons.  We choose \textbf{CTEQ6l1} \cite{cteq} parton distributions with the factorization and renormalization scales kept fixed at the parton center-of-mass energy. Initial state radiation (ISR), decay of KK-particles, showering and hadronization are also simulated with \textbf{PYTHIA}. However, \textbf{PYTHIA} implementation of mUED \cite{ElKacimi:2009zj} assumes GMD to be smaller than the KKCD (which is true for $N=6$) and hence, gravity mediated decays for heavier level-1 KK-particles are ignored. Gravity mediated decay of LKP ($\ga_1$) into a $\ga G^{\vec{n}}$-pair is considered only. However, as it has been argued, the GMD widths could be comparable (or even dominant in some parts of parameter space) with KKCD widths for $N=4~{\rm and}~2$ and hence, the GMD modes for heavier level-1 KK-particles can not be ignored. Moreover, mixing angle (Weinberg angle) between $B_\mu^1$ and $W_{3\mu}^1$ being extremely small, the LKP $\ga_1$ is essentially the level-1 excitation of $B_\mu$ and hence, can decay to both  $\ga  G^{\vec{n}}$-pair and $Z  G^{\vec{n}}$-pair. The later decay mode is ignored in the \textbf{PYTHIA} implementation of mUED. We have modified PYTHIA PYWIDTH subroutine to accommodate all possible GMD modes for all level-1 KK-particles.  For the reconstruction of physics objects (jets, leptons, photons, $\slashed E_T$ etc.) and selection of signal events, we closely follow the prescription of Ref.~\cite{ATLAS_multijet13} for dijet$+\slashed E_T$ analysis and Ref.~\cite{ATLAS_di-photon13} for di-photon$+\slashed E_T$ analysis. Jets are reconstructed with \textbf{FastJet} \cite{fastjet} implementation of anti-$k_T$ clustering algorithm \cite{antikt}. Finally, the signal cross sections for different signal regions (defined in Table~\ref{tb:multijet_cuts}~and~\ref{tb:di-photon_cuts_new}) are compared with the respective ATLAS observed 95\% CL upper limits (also shown in the same Tables). The final results are presented in Fig.~\ref{fig:exclusion} and discussed in the next section.}

\begin{figure*}[!htb]
\includegraphics[width=0.5\textwidth]{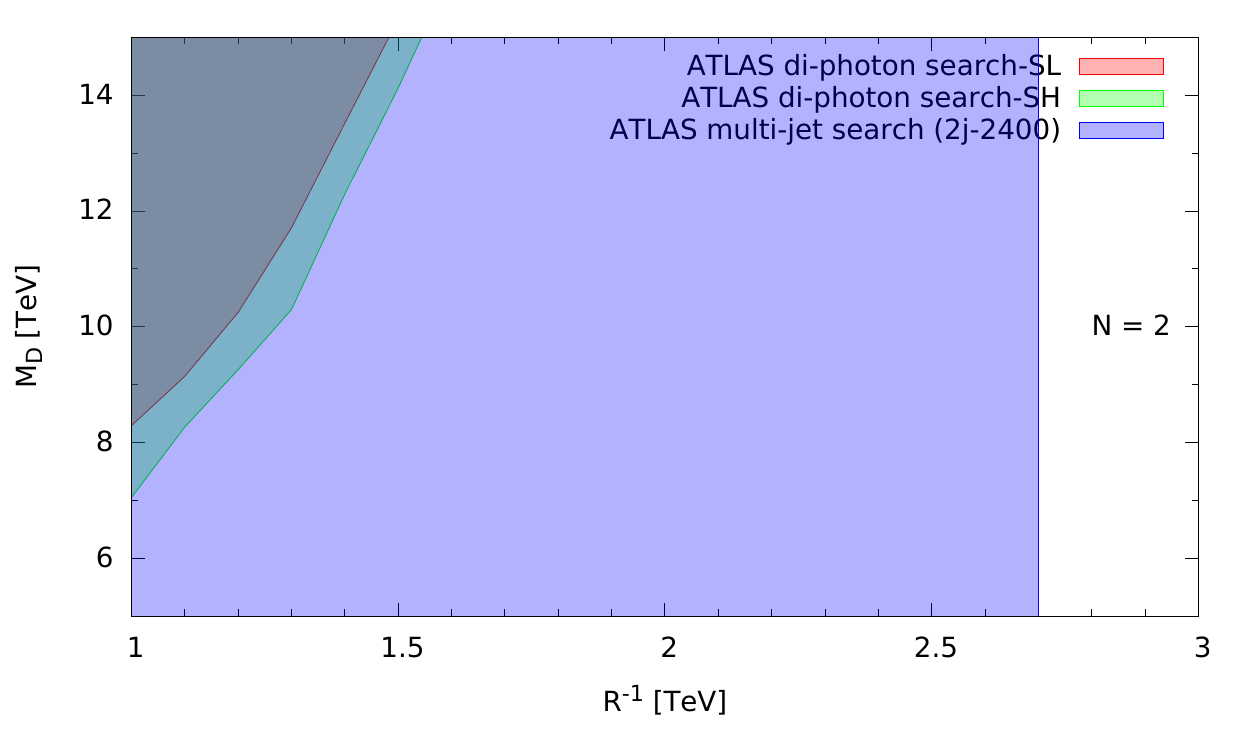}
\includegraphics[width=0.5\textwidth]{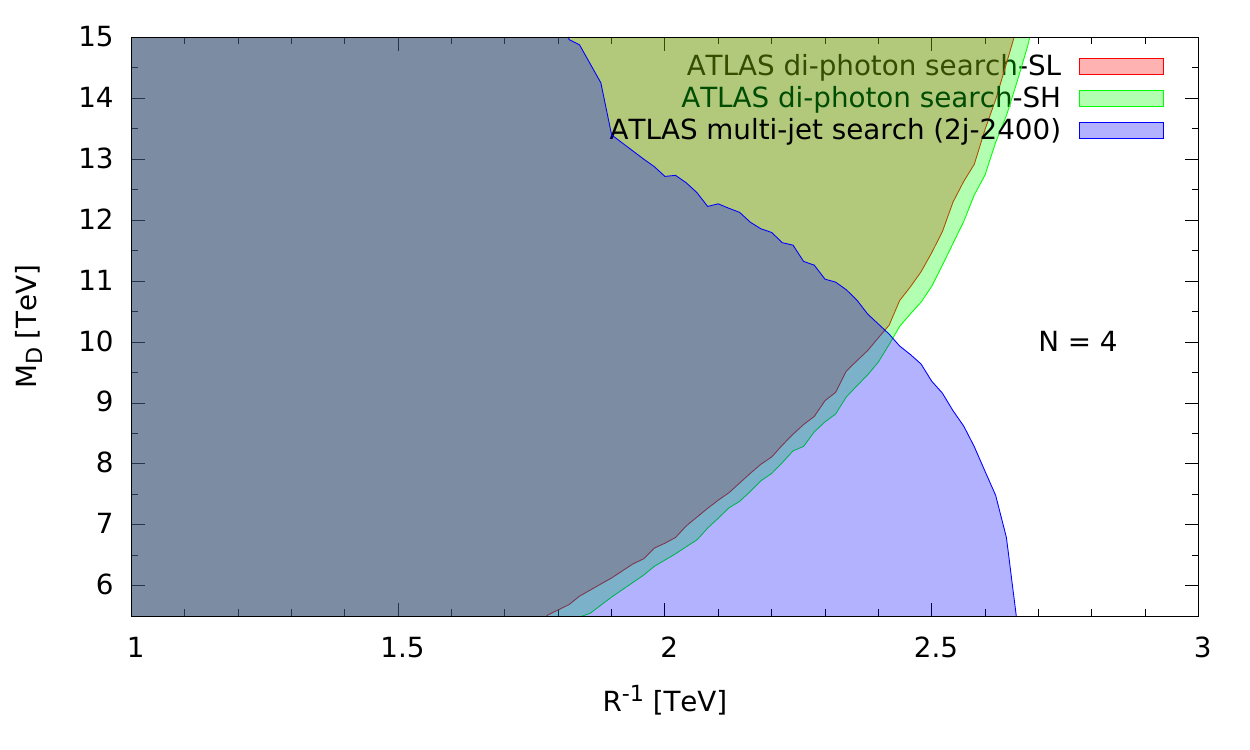}
\begin{center}
\includegraphics[width=0.5\textwidth]{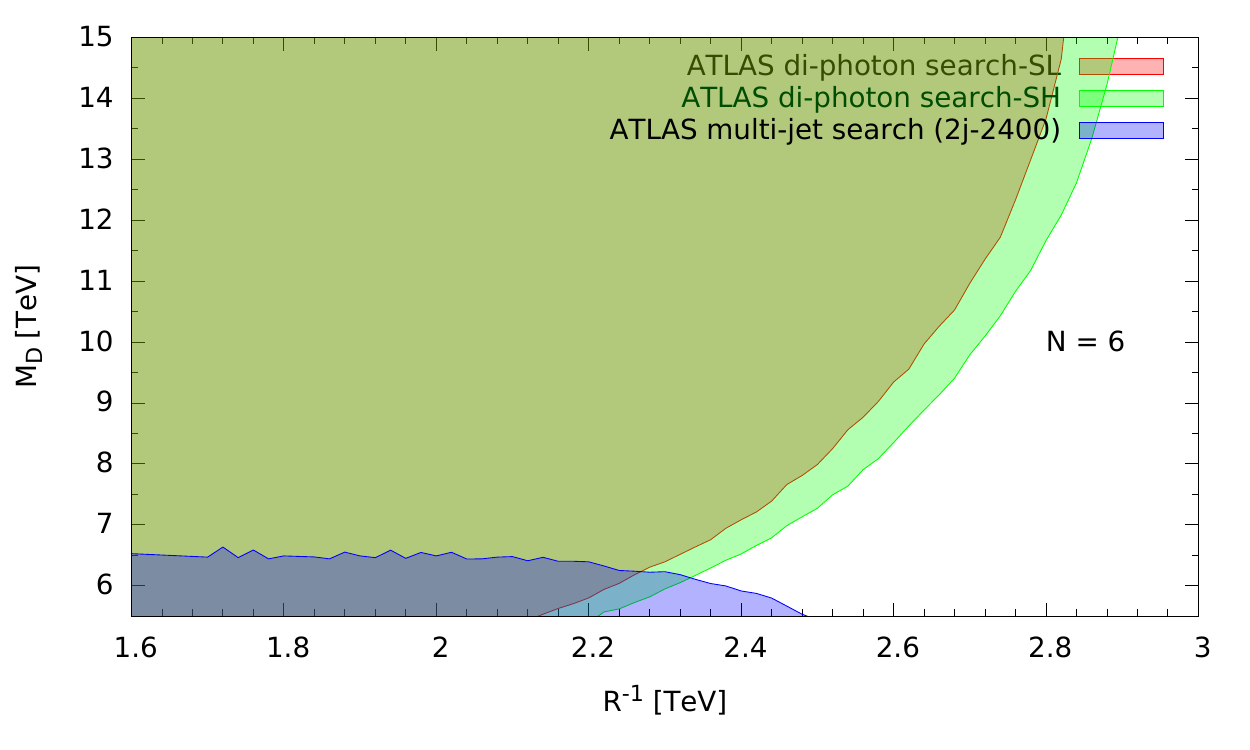}
\end{center}
\caption{The exclusion region of `fat-brane' UED model parameters $R^{-1}$ and $M_{D}$ from ATLAS multi-jet \cite{ATLAS_multijet13} (green (and blue for $N=6$)) and di-photon \cite{ATLAS_di-photon13} (red) searches for $N = 2$ (top left panel),$4$  (top right panel) and $6$ (bottom panel). $\Lambda R = 5$ is assumed throughout the analysis.}
\label{fig:exclusion}
\end{figure*}

\subsection{Bounds on 'fat brane' UED scenario}
{The collider signatures of 'fat brane' UED scenario depend on three parameters, namely the radius of small extra dimension, $R$, number of large extra dimension, $N$, and fundamental $4+N$ dimensional Planck mass, $M_D$. The radius of universal extra dimension $R$ determines the mass scale of the level-1 KK-excitations of SM particles and hence, the production cross-sections at the LHC. Whereas, $N$ and $M_D$ determine the mass splitting between the gravity excitations and hence, the density of gravity KK-states\footnote{Smaller $N$ and $M_D$ corresponds to smaller mass splitting and hence, larger density.} and strength of GMD widths. Therefore, the signal cross-sections for different ATLAS defined signal regions crucially depend on $R^{-1}$, $M_D$ and $N$. We have scanned $R^{-1}$ and $M_D$ in the range of $[1,3]$ TeV and $[5, 15]$ TeV, respectively and compared 'fat brane' UED contributions to different SRs with the ATLAS observed 95\% CL upper limits. The regions of $R^{-1}$--$M_D$ plane excluded from different LHC 13 TeV searches  are shown in Fig.~\ref{fig:exclusion} for $N=$ 2(top left panel), 4(top right panel) and 6(bottom panel).

\noindent{\it \uline{Exclusion limits for $N=2$}:} Due to smaller(larger) mass splitting between (density of) KK-gravity excitations for $N=2$, GMD widths are large and dominate over KKCD widths. As a result, after being pair produced at the LHC, KK-quarks/gluons dominantly decay into a SM quark/gluon in association with a gravity excitation and give rise to di-jet$+\slashed E_T$ signature. Fig.~\ref{fig:exclusion}(top left panel) shows that for $N=2$, the region below $R^{-1}=2.7$ TeV is excluded form ATLAS dijet$+\slashed E_T$ search (in particular, by SR 2j-2400\footnote{We have studied all the dijet signal regions defined in Table~\ref{tb:multijet_cuts}. We have also studied 3 and 4-jets signal regions defined in Ref.~\cite{ATLAS_multijet13} (but not shown in this paper). We found that strongest bounds arise from SR 2j-2400 and hence, in Fig.~\ref{fig:exclusion}, we have only presented bounds corresponding to SR 2j-2400.}). This bound is independent of $M_D \in [5, 15]$ TeV. On the other hand, ATLAS  di-photon$+\slashed E_T$ search only excluded a small part of parameter space in the large-$M_D$ and  small-$R^{-1}$ region. In particular, $R^{-1}<1000(1540)$ GeV for $M_D=7050(15000)$ GeV is excluded from  di-photon$+\slashed E_T$ results. This can be attributed to the fact that GMD(KKCD) widths decrease(increase) with increasing $M_D$($R^{-1}$) and hence, in large-$M_D$ and  small-$R^{-1}$ region, cascading of few pair produced KK-quark/gluon to LKP via KK-number conserving interactions followed by gravity mediated decay of LKP gives rise to few  di-photon$+\slashed E_T$ events. 

\noindent{\it \uline{Exclusion limits for $N=4$}:} The situation changes drastically for $N=4$ case in which KKCD widths become comparable with the GMD widths. The interplay between the strengths of GMD and KKCD resulting into dijet$+\slashed E_T$ or  di-photon$+\slashed E_T$ signatures in different parts of parameter space is clearly visible in Fig.~\ref{fig:exclusion}(top right panel).  As discussed in the previous paragraph, for low(high)-$M_D$, GMD(KKCD) dominates and hence, stringent limit arises from  dijet(di-photon)$+\slashed E_T$ search. Therefore, for $N=4$, both searches are sensitive to different (and also complementary) parts of the parameter space. In particular, we found that for $M_D=5(15)$ TeV, di-photon search excludes $R^{-1}$ below $1740(2690)$ GeV and corresponding lower limit from dijet search is $2665(1820)$ GeV. 

\noindent{\it \uline{Exclusion limits for $N=6$}:} In this case, the KKCD dominates over the GMD. Therefore, pair produced of level-1 KK-quarks/gluons decay into a pair of $\ga_1$ via cascade involving other level-1 KK-particles. Subsequent gravity mediated decay of $\ga_1$'s into photons or $Z$-bosons in association with gravity excitations gives rise to di-photon, $ZZ$ or $\gamma Z$ plus  $\slashed E_T$ signatures. We have studied di-photon$+\slashed E_T$ signature and Fig.~\ref{fig:exclusion}(bottom panel) shows that exclusion region is dominated by ATLAS di-photon$+\slashed E_T$ results. Whereas,  small part of parameter space in the low-$M_D$ region is also sensitive to dijet$+\slashed E_T$ search.  For instance, for $M_D=5(15)$ TeV, $R^{-1}<2120(2880)$ GeV region is solely excluded by di-photon$+\slashed E_T$ results. On the other hand, dijet$+\slashed E_T$ search is only sensitive for $M_D<6.5$ TeV and excludes $R^{-1}$ below 2.5 TeV.

}

\section{Conclusion and Discussion}
\label{sec:7}
{To summarize, we have studied the phenomenology of `fat-brane' UED scenario in the context of the LHC run II data. In particular, we used ATLAS searches for multi-jet$+\slashed E_T$ and di-photon$+\slashed E_T$ signatures (with $\sqrt s= 13$ TeV and $36.1$ fb$^{-1}$ integrated luminosity data) to constrain the parameter space of this model. Di-photon$+\slashed E_T$ as a signature of `fat-brane' UED scenario was previously studied by the ATLAS collaboration with $\sqrt s=7$ TeV and 3.1 pb$^{-1}$ integrated luminosity data \cite{ATLAS7tev}. The previous ATLAS analysis was done for $N=6$ and $M_D=5$ TeV and assumed 100\% branching ratio for $\ga_1 \to \gamma G^{\vec{n}}$ and neglected the gravity mediated decays of other level-1 KK-particles. In this work, we have relaxed these assumptions and performed a detailed analysis of `fat-brane' UED scenario. We found that gravity mediated decays of level-1 KK-particles are significant for $N=2$ and $4$ and hence, can not be ignored. Even for $N=6$ with low-$M_D(\sim 5~{\rm TeV})$, gravity mediated decays significantly alter the decay cascade of the level-1 KK-paticles. When the gravity mediated decays dominate over the KK-conserving decays, pair production of KK-quarks/gluons gives rise to multi-jet$+\slashed E_T$ signatures.  Depending on the parameters of the model, namely $N$, $M_D$ and $R^{-1}$, the KK-number conserving decays may also dominate over gravity-mediated decays as well resulting in di-photon$+\slashed E_T$ signature. We found that multi-jet and di-photon searches are sensitive to different (and also complementary) regions of the parameter space. For instance, the LHC 13 TeV and $36.1$ fb$^{-1}$ multi-jet(di-photon)$+\slashed E_T$ data excludes $R^{-1}$ below 2.7(2.9) TeV for $M_D=15$ TeV and $N=2(6)$. Similarly, for $M_D=5(15)$ TeV and $N=4$, a lower limit of 2.7 TeV on $R^{-1}$ arises from ATLAS multi-jet(di-photon) search. All these limits on $R^{-1}$ for different $N$ and $M_D$ are larger by a factor of 3.5 or more than the previously obtained limits in Ref.~\cite{ATLAS7tev}.

 }

\section*{Acknowledgement}
 DK thanks to the organizers of SUSY17 at Tata Institute of Fundamental Research in Mumbai, India where the initial results of the work is presented.  The work of SN  is supported in part by the US Department of Energy Grant No. de-sc 0016013. SN also would like the thank the Physics and Astronomy Department of  the Rice University (where is a visiting professor for the 2017-18 academic year), especially the High energy Physics Group, for warm hospitality
 and support.




\begin{thebibliography}{99}


\bibitem{ADD} N.~Arkani-Hamed, S.~Dimopoulous and G.~Dvali, {
  Phys.~Lett.~}{B~}{\bf 429}, 263 (1998); I.~Antoniadis,
  N.~Arkani-Hamed, S.~Dimopoulos and G.~R.~Dvali, {
  Phys.~Lett.~}{B~}{\bf 436}, 257 (1998).
\bibitem{RS} L.~Randall and R.~Sundrum, { Phys.~Rev.~Lett.~}{\bf
83}, 3370 (1999); {\em ibid} {\bf 83}, 4690 (1999).
\bibitem{antoniadis1} I.~Antoniadis,
  Phys.\ Lett.\ B {\bf 246} (1990) 377.
\bibitem{acd}
T.~Appelquist, H.~C.~Cheng and B.~A.~Dobrescu,
  Phys.\ Rev.\ D {\bf 64} (2001) 035002; \\
H.~C.~Cheng, K.~T.~Matchev and M.~Schmaltz,
  Phys.\ Rev.\  D {\bf 66} (2002) 056006.
\bibitem{relax}
  G.~Bhattacharyya, S.~K.~Majee and A.~Raychaudhuri,
  Nucl.\ Phys.\  B {\bf 793} (2008) 114.


\bibitem{Arkani-Hamed:2000hv}
  N.~Arkani-Hamed, H.~C.~Cheng, B.~A.~Dobrescu and L.~J.~Hall,
  Phys.\ Rev.\ D {\bf 62} (2000) 096006.

\bibitem{dienes} K. Dienes, E. Dudas, and T. Gherghetta; Nucl. \ Phys.\ B 
{\bf 537} (1999) 47;
  K.~R.~Dienes, E.~Dudas and T.~Gherghetta,
  Phys.\ Lett.\ B {\bf 436} (1998) 55;
  S.~Hossenfelder,
  Phys.\ Rev.\ D {\bf 70} (2004) 105003;
  G.~Bhattacharyya, A.~Datta, S.~K.~Majee and A.~Raychaudhuri,
  Nucl.\ Phys.\  B {\bf 760} (2007) 117.

\bibitem{Arkani-Hamed:1999dc}
  N.~Arkani-Hamed and M.~Schmaltz,
  Phys.\ Rev.\ D {\bf 61} (2000) 033005.
\bibitem{darkued1}
  G.~Servant and T.~M.~P.~Tait,
  Nucl.\ Phys.\  B {\bf 650}, 391 (2003).
  H.~C.~Cheng, J.~L.~Feng and K.~T.~Matchev,
  Phys.\ Rev.\ Lett.\  {\bf 89}, 211301 (2002).
  K.~Kong and K.~T.~Matchev,
  JHEP {\bf 0601}, 038 (2006);
  D.~Hooper and S.~Profumo,
  Phys.\ Rept.\  {\bf 453}, 29 (2007);
  T.~Flacke, D.~W.~Kang, K.~Kong, G.~Mohlabeng and S.~C.~Park,
  JHEP {\bf 1704}, 041 (2017);
  G.~Belanger, M.~Kakizaki and A.~Pukhov,
  JCAP {\bf 1102}, 009 (2011);
  M.~Kakizaki, S.~Matsumoto and M.~Senami,
  Phys.\ Rev.\ D {\bf 74}, 023504 (2006);
  F.~Burnell and G.~D.~Kribs,
  Phys.\ Rev.\ D {\bf 73}, 015001 (2006);
  Y.~Ishigure, M.~Kakizaki and A.~Santa,
  arXiv:1611.06760 [hep-ph];
  J.~M.~Cornell, S.~Profumo and W.~Shepherd,
  Phys.\ Rev.\ D {\bf 89}, no. 5, 056005 (2014).



\bibitem{darkued2}
  B.~A.~Dobrescu, D.~Hooper, K.~Kong and R.~Mahbubani,
  JCAP {\bf 0710}, 012 (2007);
  M.~T.~Arun, D.~Choudhury and D.~Sachdeva,
  arXiv:1805.01642 [hep-ph].

\bibitem{2UED_fg}
  B.~A.~Dobrescu and E.~Poppitz,
  Phys.\ Rev.\ Lett.\  {\bf 87}, 031801 (2001).
\bibitem{2UED_proton} 
  T.~Appelquist, B.~A.~Dobrescu, E.~Ponton and H.~U.~Yee,
  Phys.\ Rev.\ Lett.\  {\bf 87}, 181802 (2001).


\bibitem{collider_ED} 
  A.~Datta, G.~L.~Kane and M.~Toharia,
  hep-ph/0510204;
  A.~J.~Barr,
  JHEP {\bf 0602}, 042 (2006);
  B.~Bhattacherjee and A.~Kundu,
  Phys.\ Lett.\ B {\bf 653}, 300 (2007);
  P.~Bandyopadhyay, B.~Bhattacherjee and A.~Datta,
  JHEP {\bf 1003}, 048 (2010);
  D.~Choudhury, A.~Datta and K.~Ghosh,
  JHEP {\bf 1008}, 051 (2010);
  K.~Kong, K.~Matchev and G.~Servant,
  In *Bertone, G. (ed.): Particle dark matter* 306-324;
  B.~Bhattacherjee and K.~Ghosh,
  Phys.\ Rev.\ D {\bf 83}, 034003 (2011);
  A.~Datta, A.~Datta and S.~Poddar,
  Phys.\ Lett.\ B {\bf 712}, 219 (2012);
  A.~Datta, K.~Kong and K.~T.~Matchev,
  Phys.\ Rev.\ D {\bf 72}, 096006 (2005)
  Erratum: [Phys.\ Rev.\ D {\bf 72}, 119901 (2005)];
  T.~G.~Rizzo,
  Phys.\ Rev.\ D {\bf 64}, 095010 (2001);
  H.~C.~Cheng,
  Int.\ J.\ Mod.\ Phys.\ A {\bf 18}, 2779 (2003);
  A.~Muck, A.~Pilaftsis and R.~Ruckl,
  Nucl.\ Phys.\ B {\bf 687}, 55 (2004);
  M.~Battaglia, A.~Datta, A.~De Roeck, K.~Kong and K.~T.~Matchev,
  JHEP {\bf 0507}, 033 (2005);
  U.~K.~Dey and T.~Jha,
  Phys.\ Rev.\ D {\bf 94}, no. 5, 056011 (2016);
  U.~K.~Dey and A.~Raychaudhuri,
  Nucl.\ Phys.\ B {\bf 893}, 408 (2015);
  T.~Flacke, K.~Kong and S.~C.~Park,
  Mod.\ Phys.\ Lett.\ A {\bf 30}, no. 05, 1530003 (2015);
  D.~Kim, Y.~Oh and S.~C.~Park,
  J.\ Korean Phys.\ Soc.\  {\bf 67}, 1137 (2015);
  H.~Murayama, M.~M.~Nojiri and K.~Tobioka,
  Phys.\ Rev.\ D {\bf 84}, 094015 (2011).




\bibitem{ED_LHC} 
  S.~Chatrchyan {\it et al.} [CMS Collaboration],
  Phys.\ Rev.\ Lett.\  {\bf 108}, 111801 (2012);
  S.~Chatrchyan {\it et al.} [CMS Collaboration],
  JHEP {\bf 1105}, 093 (2011);
  S.~Chatrchyan {\it et al.} [CMS Collaboration],
  Phys.\ Rev.\ D {\bf 87}, no. 11, 114015 (2013);
  V.~Khachatryan {\it et al.} [CMS Collaboration],
  Eur.\ Phys.\ J.\ C {\bf 75}, no. 5, 235 (2015);
  G.~Aad {\it et al.} [ATLAS Collaboration],
  JHEP {\bf 1603}, 026 (2016);
  G.~Aad {\it et al.} [ATLAS Collaboration],
  Phys.\ Rev.\ D {\bf 92}, no. 3, 032004 (2015);
  G.~Aad {\it et al.} [ATLAS Collaboration],
  JHEP {\bf 1508}, 148 (2015);
  G.~Aad {\it et al.} [ATLAS Collaboration],
  JHEP {\bf 1504}, 116 (2015);
  L.~Morvaj,
  CERN-THESIS-2014-284;
  D.~Choudhury and K.~Ghosh,
  Phys.\ Lett.\ B {\bf 763}, 155 (2016);
  J.~Beuria, A.~Datta, D.~Debnath and K.~T.~Matchev,
  Comput.\ Phys.\ Commun.\  {\bf 226}, 187 (2018);
  L.~Edelhäuser, T.~Flacke and M.~Krämer,
  JHEP {\bf 1308}, 091 (2013);
  G.~Cacciapaglia, A.~Deandrea, J.~Ellis, J.~Marrouche and L.~Panizzi,
  Phys.\ Rev.\ D {\bf 87}, no. 7, 075006 (2013);
  G.~Servant,
  Mod.\ Phys.\ Lett.\ A {\bf 30}, no. 15, 1540011 (2015).
  K.~Ghosh, D.~Karabacak and S.~Nandi,
  JHEP {\bf 1409}, 076 (2014).
 
\bibitem{coll2ued}
  G.~Burdman, B.~A.~Dobrescu and E.~Ponton,
  Phys.\ Rev.\  D {\bf 74}, 075008 (2006);
  B.~A.~Dobrescu, K.~Kong and R.~Mahbubani,
  JHEP {\bf 0707}, 006 (2007);
  A.~Freitas and K.~Kong,
  JHEP {\bf 0802}, 068 (2008);
  K.~Ghosh and A.~Datta,
  Nucl.\ Phys.\  B {\bf 800}, 109 (2008);
  K.~Ghosh and A.~Datta,
  Phys.\ Lett.\  B {\bf 665}, 369 (2008);
  K.~Ghosh,
  JHEP {\bf 0904}, 049 (2009);
  D.~Choudhury, A.~Datta, D.~K.~Ghosh and K.~Ghosh,
 JHEP {\bf 1204}, 057 (2012);
  G.~Burdman, O.~J.~P.~Eboli and D.~Spehler,
  Phys.\ Rev.\ D {\bf 94}, no. 9, 095004 (2016).

\bibitem{NPB550}
A.~Donini,~S.~Rigolin;
{ Nucl. Phys.} {\bf B550}, 59 (1999);
I. Antoniadis, K. Benakli, M. Quiros;
{Phys. Lett. B} {\bf 460}, 176 (1999).
\bibitem{PLB482}
A. De Rujula , A. Donini, M. B. Gavela, S. Rigolin,
Phys. Lett. B {\bf 482}, 195 (2000);
D. A. Dicus, C. D. McMullen, S. Nandi,
Phys. Rev. D {\bf 65}, 076007 (2002);
C. Macesanu, C. D. McMullen, S. Nandi, 
Phys. Lett. B {\bf 546}, 253 (2002); 
C. Macesanu, S. Nandi, C. M. Rujoiu, 
Phys. Rev. D {\bf 73}, 076001 (2006);
  C.~Macesanu, S.~Nandi and M.~Rujoiu,
  Phys.\ Rev.\ D {\bf 71}, 036003 (2005);
  E.~Gabrielli and B.~Mele,
  Nucl.\ Phys.\ B {\bf 647}, 319 (2002).
\bibitem{Macesanu16} 
  C.~Macesanu, A.~Mitov and S.~Nandi,
  Phys.\ Rev.\ D {\bf 68}, 084008 (2003).




\bibitem{radi_matchev}
  H.~C.~Cheng, K.~T.~Matchev and M.~Schmaltz,
  Phys.\ Rev.\  D {\bf 66} (2002) 036005.

\bibitem{UED_VS2} 
  A.~Datta and S.~Raychaudhuri,
  Phys.\ Rev.\ D {\bf 87}, no. 3, 035018 (2013).
\bibitem{Datta:2013xwa} 
  A.~Datta, A.~Patra and S.~Raychaudhuri,
  Phys.\ Rev.\ D {\bf 89}, no. 9, 093008 (2014).
\bibitem{Ghosh:2012zc} 
  K.~Ghosh and K.~Huitu,
  JHEP {\bf 1206}, 042 (2012).
\bibitem{ATLAS_multijet13}
The ATLAS collaboration [ATLAS Collaboration],
  ATLAS-CONF-2017-022.


\bibitem{ATLAS_di-photon13}
M.~Aaboud {\it et al.} [ATLAS Collaboration],
  arXiv:1802.03158 [hep-ex].
\bibitem{antikt} 
  M.~Cacciari, G.~P.~Salam and G.~Soyez,
  JHEP {\bf 0804}, 063 (2008).

\bibitem{PYTHIA} 
T.~Sjostrand{\it et al.},
JHEP {\bf 0605}, 026 (2006).
\bibitem{ElKacimi:2009zj} 
  M.~ElKacimi, D.~Goujdami, H.~Przysiezniak and P.~Z.~Skands,
  Comput.\ Phys.\ Commun.\  {\bf 181}, 122 (2010).
\bibitem{cteq} 
  J.~Pumplin, D.~R.~Stump, J.~Huston, H.~L.~Lai, P.~M.~Nadolsky and W.~K.~Tung,
  JHEP {\bf 0207}, 012 (2002).
\bibitem{fastjet} 
  M.~Cacciari, G.~P.~Salam and G.~Soyez,
  Eur.\ Phys.\ J.\ C {\bf 72}, 1896 (2012).
  

\bibitem{ATLAS7tev} 
  G.~Aad {\it et al.}  [ATLAS Collaboration],
  Phys.\ Rev.\ Lett.\  {\bf 106}, 121803 (2011).




\end{thebibliography}


\end{document}